\documentclass[%
prl,
reprint,
superscriptaddress,
preprintnumbers,
nofootinbib,
amsmath,amssymb,
aps,
]{revtex4-1}
\pdfoutput=1
\usepackage{hyperref}
\usepackage{graphicx}

\usepackage[textwidth=1.25cm,textsize=footnotesize,
]{todonotes}

\usepackage[utf8]{inputenc}

\usepackage{color}
\usepackage[normalem]{ulem}

\newcommand{\be}{\begin{equation}}
\newcommand{\ee}{\end{equation}}
\newcommand{\ba}{\begin{eqnarray}}
\newcommand{\ea}{\end{eqnarray}}

\renewcommand{\l}{\left(}
\renewcommand{\r}{\right)}

\def\aap{Astronomy and Astrophys.}

\def\prd{Phys. Rev. D}

\def\srg{\emph{SRG}}

\def\nustar{\emph{NuSTAR}}

\begin{document}
\title{All-sky limits on sterile neutrino galactic dark matter\\ obtained with SRG/ART-XC after two years of operations
}

\author{E.\,I.\,Zakharov}
\email[]{ezakharov@cosmos.ru}
\affiliation{%
  Space Research Institute of the Russian Academy of Sciences, Moscow 117997, Russia}
\affiliation{%
  National Research University Higher School of Economics, Moscow 101000, Russia}
\affiliation{%
  Institute for Nuclear Research of the Russian Academy of Sciences, Moscow 117312, Russia}
  
\author{V.\,V.\,Barinov}
\email[]{barinov.vvl@gmail.com}
\affiliation{%
  Physics Department, M. V. Lomonosov Moscow State University, Leninskie Gory, Moscow 119991, Russia}
\affiliation{%
  Institute for Nuclear Research of the Russian Academy of Sciences, Moscow 117312, Russia}
\author{R.\,A.\,Burenin}
\email[]{rodion@hea.iki.rssi.ru}
\affiliation{%
  Space Research Institute of the Russian Academy of Sciences, Moscow 117997, Russia}
\affiliation{%
  National Research University Higher School of Economics, Moscow 101000, Russia}

\author{D.\,S.\,Gorbunov}
\email[]{gorby@ms2.inr.ac.ru}
\affiliation{%
  Institute for Nuclear Research of the Russian Academy of Sciences, Moscow 117312, Russia}
\affiliation{%
  Moscow Institute of Physics and Technology, Dolgoprudny 141700, Russia}
\author{R.\,A.\,Krivonos}
\email[]{krivonos@cosmos.ru}
\affiliation{%
  Space Research Institute of the Russian Academy of Sciences, Moscow 117997, Russia}
\affiliation{%
  Institute for Nuclear Research of the Russian Academy of Sciences, Moscow 117312, Russia}

\author{A.\,Yu.\,Tkachenko}
\affiliation{%
  Space Research Institute of the Russian Academy of Sciences, Moscow 117997, Russia}

\author{V.\,A.\,Arefiev}
\affiliation{%
  Space Research Institute of the Russian Academy of Sciences, Moscow 117997, Russia}

\author{E.\,V.\,Filippova}
\affiliation{%
  Space Research Institute of the Russian Academy of Sciences, Moscow 117997, Russia}

\author{S.\,A.\,Grebenev}
\affiliation{%
  Space Research Institute of the Russian Academy of Sciences, Moscow 117997, Russia}
 
\author{A.\,A.\,Lutovinov}
\affiliation{%
  Space Research Institute of the Russian Academy of Sciences, Moscow 117997, Russia}

\author{I.\,A.\,Mereminsky}
\affiliation{%
  Space Research Institute of the Russian Academy of Sciences, Moscow 117997, Russia}

\author{S.\,Yu.\,Sazonov}
\affiliation{%
  Space Research Institute of the Russian Academy of Sciences, Moscow 117997, Russia}

\author{A.\,N.\,Semena}
\affiliation{%
  Space Research Institute of the Russian Academy of Sciences, Moscow 117997, Russia}

\author{A.\,E.\,Shtykovsky}
\affiliation{%
  Space Research Institute of the Russian Academy of Sciences, Moscow 117997, Russia}
\affiliation{%
  National Research University Higher School of Economics, Moscow 101000, Russia}

\author{R.\,A.\,Sunyaev}
\affiliation{%
  Space Research Institute of the Russian Academy of Sciences, Moscow 117997, Russia}

\bigskip
\preprint{INR-TH-2023-002}

\begin{abstract}
Dark matter sterile neutrinos radiatively decay in the Milky Way, which can be tested with searches for almost monochromatic photons in the x-ray cosmic spectrum. We analyse the data of SRG/ART-XC telescope operated for two years in the all-sky survey mode. With no significant hints in  the Galactic diffuse x-ray spectrum we explore models with sterile neutrino masses in 12-40\,keV range and exclude corresponding regions of sterile-active neutrino mixing.   
\end{abstract}


\maketitle

{\it 1. Introduction.} Incompleteness of the Standard Model (SM) of particle physics is widely recognized, since it fails to explain neutrino oscillations, baryon asymmetry of the Universe, dark matter (DM) phenomena, etc. Though many extensions have been suggested in literature, those of a special interest address not just one, but two or more issues at once. 

A well-known example of this kind is provided by SM extensions with {\it sterile neutrinos}. They are massive fermions, a singlet with respect to the SM gauge group, which can mix with SM neutrinos through the mass terms\,\cite{Schechter:1980gr}. This mixing makes the SM (or {\it active}) neutrinos massive and explains the neutrino oscillations (for brief reviews see \cite{Volkas:2001zb,Gorbunov:2014efa}). At the same time this mixing induces interaction between the sterile neutrinos and SM particles that, depending on the model parameters, allows for generating the required amount of baryon asymmetry in the early Universe via leptogenesis and/or constructing the  main dark matter component from sterile neutrinos, see e.g.\,\cite{Boyarsky:2009ix,Drewes:2013gca}. In the latter case the mixing makes this dark matter unstable. To be viable but sufficiently long-lived, i.e. survive till the present cosmological epoch, sterile neutrino $\nu_s$ must be light, with mass $m_s$ in keV range, see e.g.\,\cite{Abazajian:2017tcc,Dasgupta:2021ies}. However, it is still unstable, and can decay radiatively, i.e. into an active neutrino and photon, 
\begin{equation}
  \label{decays}
\nu_{s} \rightarrow \nu_{e, \mu, \tau} + \gamma
\end{equation}
with decay rate\,\cite{PhysRevD.25.766,Barger:1995ty}
\begin{align}\label{eq:neutrino_width}
\nonumber\Gamma_{\gamma} &= \frac{9}{1024} \frac{\alpha}{\pi^4} G_F^2 m_s^5\sin^22\theta\\
&= 1.36 \times 10^{-22}\left(\frac{m_s}{1 \text{keV}}\right)^5 \sin^22\theta\hspace{0.25cm} \text{s}^{-1},
\end{align}
where $\theta$ is a sterile-active neutrino mixing angle, $G_F$ is the Fermi constant and $\alpha$ is the fine-structure constant. The radiative decay \eqref{decays} yields a very special signature of the sterile neutrino dark matter; almost monochromatic x-rays of energy $E\approx m_s/2$\,\cite{Abazajian:2001vt}. 

\begin{figure}[!ht]
    \centering
    \includegraphics[width=\columnwidth]{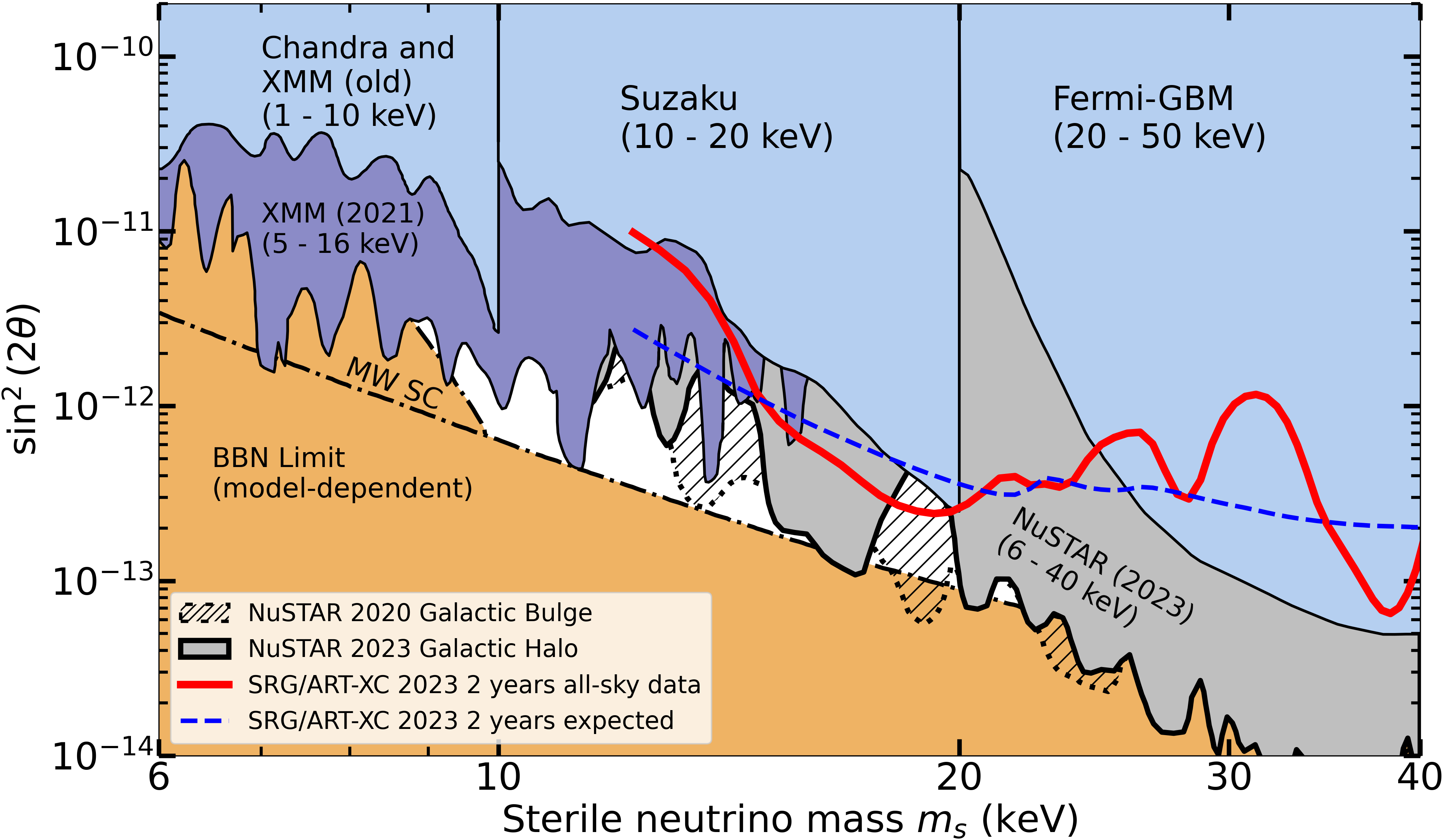}
    \caption{Upper limits on mixing (95\%\,C.L.) from 2 years of data taking at ART-XC in the survey mode are depicted with red solid line. The blue dashed line refers to the expected 2-year sensitivity of ART-XC. Note, that the expected sensitivity does not contain statistical fluctuations observed in real data. Black solid line with gray shading show limits recently obtained with the {\nustar} in the Galactic halo \cite{Roach:2022lgo}. And black dotted line with hatching show limits obtained in Galactic bulge \cite{2020PhRvD.101j3011R} with {\nustar} (see our comment in Sec. 4). The white region is remaining allowed area for sterile neutrinos parameters.  The previous upper limits: Fermi-GBM \cite{FermiGBM}, Suzaku \cite{Suzaku}, XMM 2021 \cite{XMM2021} and Chandra and XMM (old) \cite{CXO_Andromeda, XMM_Draco, CXO_Draco, Watson:2006qb}. And the previous lower limits: Milky Way satellite counts (MW SC) \cite{MWSC} and big bang nucleosynthesis (BBN) limit \cite{BBN1, BBN2}.}
    \label{fig:results}
\end{figure}

In this paper we perform searches for this signature in two year data of ART-XC telescope operated in all-sky survey mode. We find no solid evidences for a monochromatic line to be associated with decay \eqref{decays} of sterile neutrinos forming the dark matter halo of our  Milky Way (MW) Galaxy. As a result we place upper limits on the mixing angle $\theta$ at the sterile neutrino masses $m_s$ from 12\,keV to 40\,keV, which are presented in Fig.\,\ref{fig:results}, together with most relevant constraints from previous searches. 
\newpage

{\it 2. X-ray all sky spectrum.} 
The {\it Mikhail Pavlinsky\/} Astronomical Roentgen Telescope --  x-ray concentrator (ART-XC) \cite{Pavlinsky:2021ynp, 2018ExA....45..315P, 2019ExA....47....1P, 2019ExA....48..233P} is one of two Wolter type-I x-ray telescopes with grazing incidence optics on board the Spectrum-Roentgen-Gamma  (\srg) observatory \cite{Sunyaev:2021bln}. Launched in summer 2019 from the Baikonur Cosmodrome, the observatory arrived at the L2 region and after various tests and proper adjustments of its trajectory started its main scientific program operating in the all-sky survey mode. We use the data collected by ART-XC from December 12th, 2019 to December 19th, 2021, that forms four complete full-sky maps in x-rays of the energy range 6-20\,keV, to search for possible traces of the radiative decay \eqref{decays} assuming that the sterile neutrinos compose the entire dark matter component of the Milky Way.  The expected signal associated with the dark matter must be observed at one and the same energy range (where the width refers to the telescope energy resolution) and its intensity must vary over the sky reflecting the galaxy profile of the dark matter component. It has been shown in Ref.\,\cite{Barinov:2020hiq}, that investigations of the x-ray spectrum from the central part of the MW (a cone with a half-apex angle of $60^\circ$) gives us an opportunity to test previously unexplored region of the sterile neutrino model parameters.  

\begin{figure}[tb]
    \centering
    \includegraphics[width=\columnwidth]{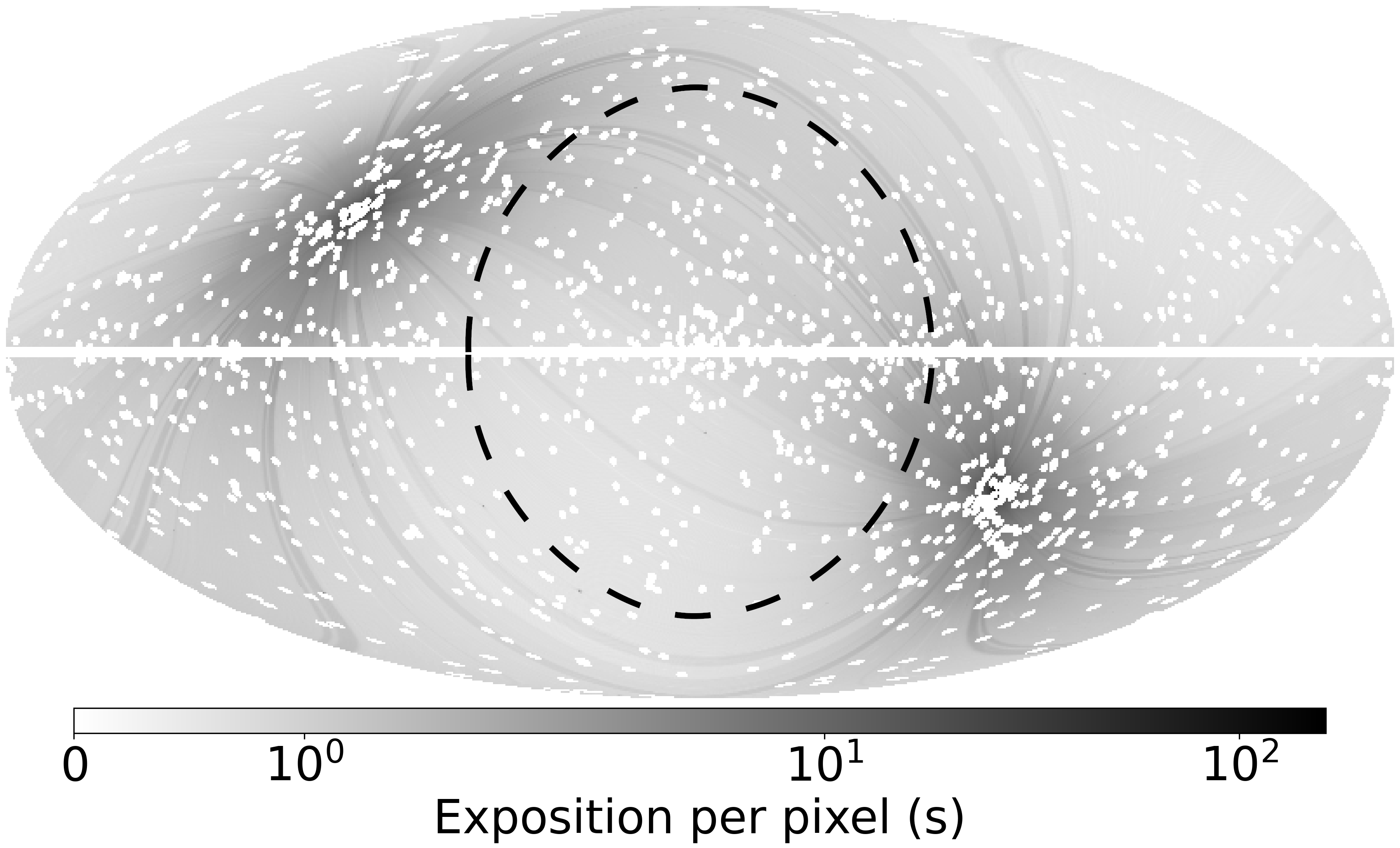}
    \caption{The two-year all-sky exposure $\epsilon(\phi,\psi)$ of ART-XC used in this analysis. The regions excluded from our analysis are shown in white. Black dashed circle denotes the border between the region I (inner area, $\Omega^I \approx 9040$ deg$^2$) and region II (outer area, $\Omega^{II}\approx 27600$ deg$^2$). Regions I and II comprise exposure time of $T^I \approx 10.6$\,Ms and $T^{II}\approx 42.5$\,Ms, respectively.}
    \label{fig:exposure}
\end{figure}

\begin{figure}[tb]
    \centering
    \includegraphics[width=\columnwidth]{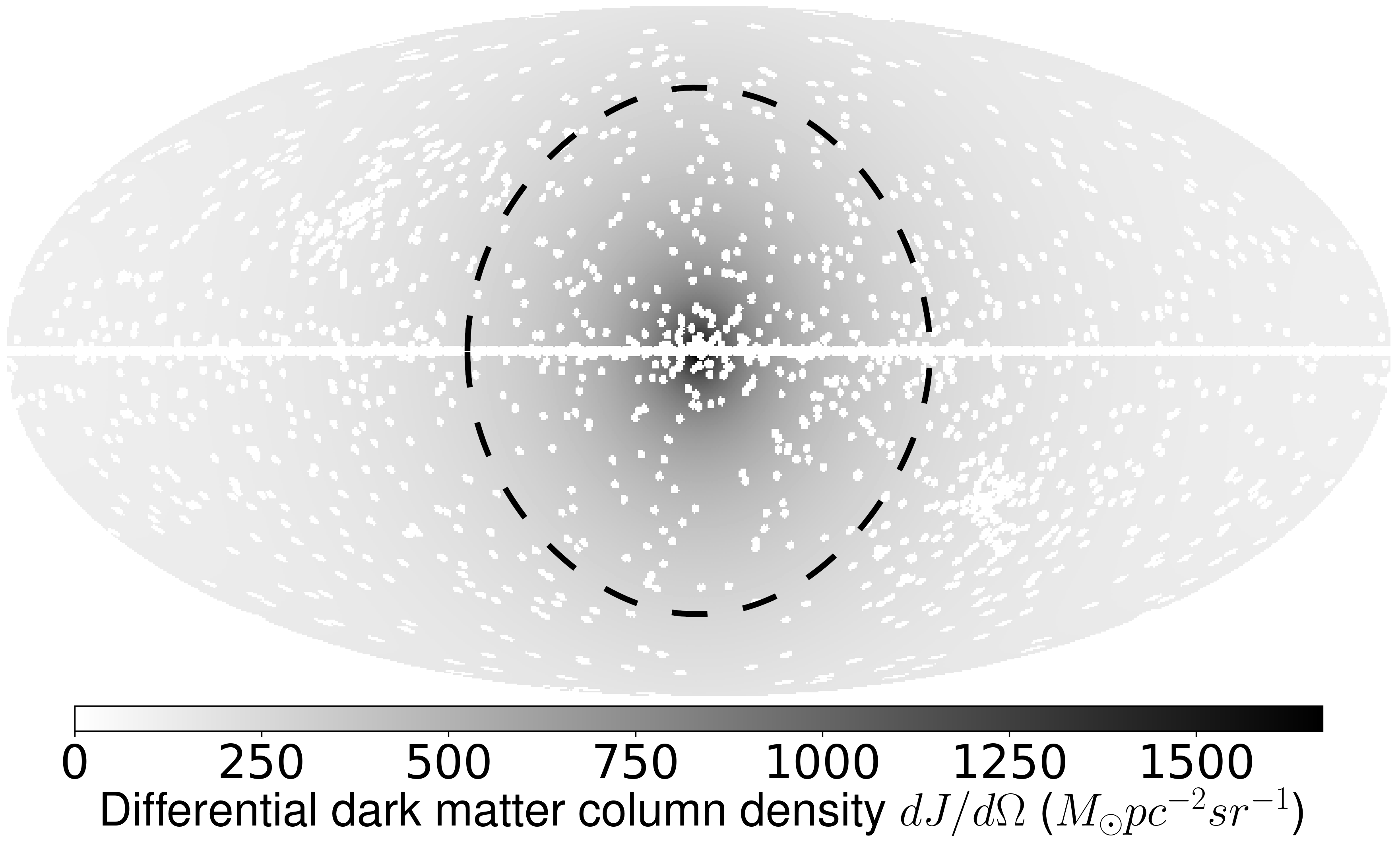}
    \caption{Differential column density for NFW profile of the MW halo.}
    \label{fig:DM}
\end{figure}

We use the diffuse part of the x-ray flux for this study, and so exclude the region of Galactic disc [galactic latitude $|b|<1^\circ$,  which corresponds to the characteristic size of the Galactic Ridge x-ray emission (GRXE) \cite{Revnivtsev:2005rj}] and circles of radius $1^\circ$ (which corresponds to the maximum offset angle for singly reflected photons) centered at the positions of all the x-ray sources from the catalog of sources detected during the two years of SRG/ART-XC  operations in the surveying mode. This catalog includes 867 sources from the first year catalog \cite{Pavlinsky:2021jsk} and additional 642 sources detected in combined data after the second year of observations.  Hereafter, neither the experimental data nor the expected signal associated with these excluded parts of the sky, painted black in Figs.\,\ref{fig:exposure} and \ref{fig:DM},    
are considered. 

We further refine the strategy of\, \cite{Barinov:2020hiq} by exploiting the survey operation mode to get rid of the possible time-dependent isotropic background. The main contribution to the ART-XC background comes from the low-energy cosmic rays, which are  expected to be isotropic and hence their flux in a given direction is proportional to the local integrated exposure presented in Fig.\,\ref{fig:exposure}. We also assume that in our energy range (from 6 keV to 20 keV) there are only two sources of diffuse x-ray emission: cosmic x-ray background (CXB) and GRXE. The CXB is fairly uniform, and the GRXE has been cut out.
Hence, to be safe from the short-term fluctuations of the detector parameters (on long-term intervals all the detector parameters are very stable which is observed through the periodical monitoring) one would  investigate the possible dark matter contribution to the difference of the normalized x-ray fluxes coming from the central part of the MW and from the opposite direction. In the survey mode the telescope rotates, observing the opposite directions each 4 hours and presumably being at the same internal operational conditions. However, this procedure would use only part of the collected data. To use the entire statistics and suppress the time-dependent isotropic background we adopt the following strategy. 

We split the sky into two regions -- the central part of MW limited by 
a cone with half apex angle of $60^\circ$ and the rest part -- see the dashed white circle along the border of the two regions in Figs.\,\ref{fig:exposure} and \ref{fig:DM}. In each region we count all highly energetic and presumably background events with photon energy in the interval  40-120\,keV which were definitely induced by the cosmic rays and find the background count ratio $\lambda=0.24960\pm0.00005$ between regions I and II. Note, that this ratio is very close to the corresponding ratio ${\sim}0.25$ of the integrated exposures for the two regions (Fig.\,\ref{fig:exposure}).  

\begin{figure}[tb]
    \centering
    \includegraphics[width=\columnwidth]{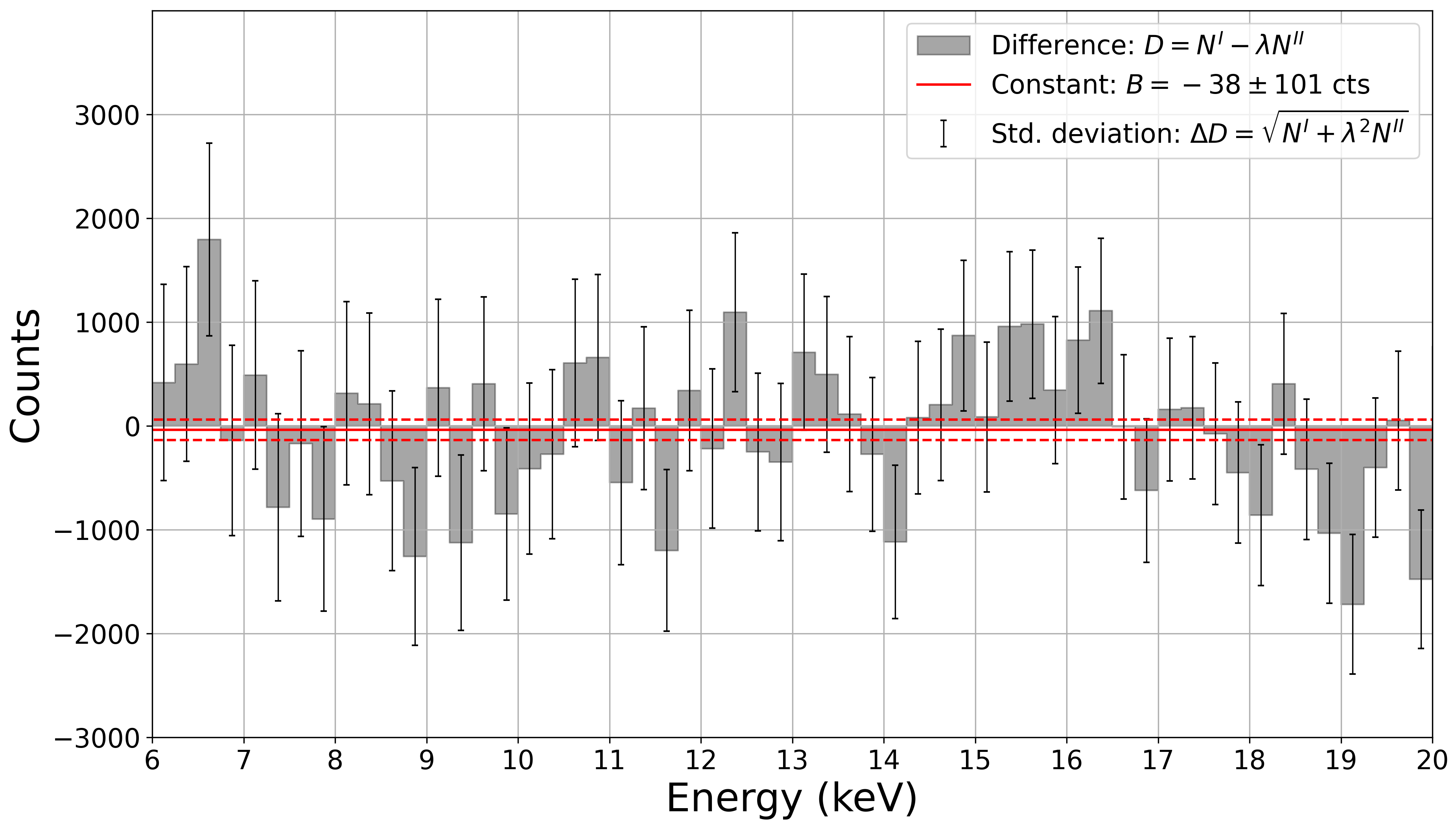}
    \caption{Differential count number $D_i$. Gray shading indicates the residuals, vertical lines show the error bars \eqref{stat-errors}. The red solid line is the fit to the residuals by constant $B$ and the red dashed lines indicate $\pm1\sigma$ confidence intervals.}
    \label{fig:residuals}
\end{figure}

Then we turn to the data in the working range of photon energies 6-20\,keV. We divide the energy range into $N^b=56$ equal energy bins each of $\Delta E = 0.25$\,keV in size. Considering events in each energy bin $i$ we subtract from the number collected in the first region $N_i^I$ that in the second region $N_i^{II}$ multiplied by $\lambda$. Without any fluctuations, this subtraction, 
\[
D_i\equiv N_i^I-\lambda N_i^{II}\,,
\] 
would average out the cosmic ray background in each energy bin to $D_i\approx 0$, while the expected signal from dark matter decay, being anisotropic over the sky, would  show up in the energy bins around energy $m_s/2$. 
 
The differential count numbers $D_i$, obtained for regions I and II, are  presented in Fig.\,\ref{fig:residuals}. 
Introducing for the residual counts in each energy bin an error of 
\begin{equation}
\label{stat-errors}
\Delta D_i \equiv \sqrt{N^I_i+\lambda^2 N^{II}_i} \approx \sqrt{1+\lambda}\, \sqrt{N^I_i}
\end{equation}
due to pure statistical independent fluctuations of the events in the first and second regions, 
we fit these residuals by constant. In each energy bin it equals (see red line in Fig.\,\ref{fig:residuals})   
\[
B\pm\Delta B=-38\pm101\,\text{cts}\,,
\]
and gives a fit with $\chi^2/d.o.f.=0.91$, which is acceptable. Now no bin deviates for more than $2\sigma$, as expected for completely similar spectra in both two regions in the sky. The  subtraction of this constant background from the data does not noticeably affect our chances in searches for the decaying dark matter, since its signature is a peak, i.e., a feature concentrated at a particular energy. We checked that changing the $B$ constant by $\pm1\sigma$ leads to ${\sim}20\%$ change in the resulting upper limits, which is significantly less than statistical fluctuations in real data (see Fig.\ref{fig:results}).

{\it 3. Data analysis.} 
The differential flux from decaying dark matter in a given direction is determined by its column density $dJ=d\Omega\int\!\text{d} l\, \rho_{DM}(\vec l)$  along the line of sight. 
Hence the signal flux expected from the central part of the MW equals
\begin{equation}\label{eq:gamma_flux_th_signal}
F^I \! = \!
\frac{\Gamma_{\gamma}}{4\pi m_s}\times\hat J^I,
\end{equation}
where the column density is corrected for the ART-XC exposure map 
\begin{equation}\label{eq:Jfactor}
\hat J^{I} \! = \!
\int_{0}^{2\pi}
\!\!\!
\int_{0}^{60^\circ}
\!\!\!\!\!
\int_{0}^{R+R_{vir}} 
\!\!\!\!\!\!\!\!\!\!\!\!\!\!\!
\rho_{\text{\tiny DM}}(r(l,\! \psi))\,\tilde\epsilon^I(\phi,\psi)\,\sin\!\psi\,\text{d}\phi\,\text{d}\psi\,\text{d}l.
\end{equation}
Here the distance from the MW center is $r(l, \psi) = \sqrt{R^2 - 2R l\cos\psi + l^2} $, $\psi$ is the angle from the direction to the MW center, $ R=8$\,kpc is the distance from the observer to the MW center, the virial radius is $R_{vir}=200$\,kpc\,\cite{Dehnen:2006cm}, $l$ is the distance along the line of sight and $\tilde\epsilon(\phi,\psi)$ is the local exposure for the two-year operation, see Fig.\,\ref{fig:exposure}, divided by the total observation time of the given region, $\tilde \epsilon^I(\phi,\psi)\equiv \epsilon(\phi,\psi)/T^I$. The signal flux from the second region, $F^{II}$ is given by \eqref{eq:gamma_flux_th_signal} upon replacing the integration range $(0,60^\circ)\to (60^\circ,180^\circ)$. In this analysis we use the standard Navarro-Frenk-White (NFW) profile $\rho_{\text{\tiny DM}}(r) = \rho_{s}/\left(r/r_s\right)\left(1 +
r/r_s\right)^{2}$ \cite{Navarro:1995iw} with $\rho_{s} = 10.5\times
10^{-3}\text{M}_{\bigodot} \text{pc}^{-3}$ and $r_s = 20$\,kpc~\cite{2017PhRvD..95l3002P}. Note that the result of integration in~\eqref{eq:Jfactor} is almost insensitive to the choice of the upper limit on $l$ if the upper limit is $R_{vir}$ or slightly larger. For example, the result changes by less than 0.2\% when the upper limit is changed from $R + R_{vir}$ to 0.5 Mpc. Figure\,\ref{fig:DM} illustrates the predicted signal distribution over the sky. 

\begin{figure}[tb]
    \centering
    \includegraphics[width=\columnwidth]{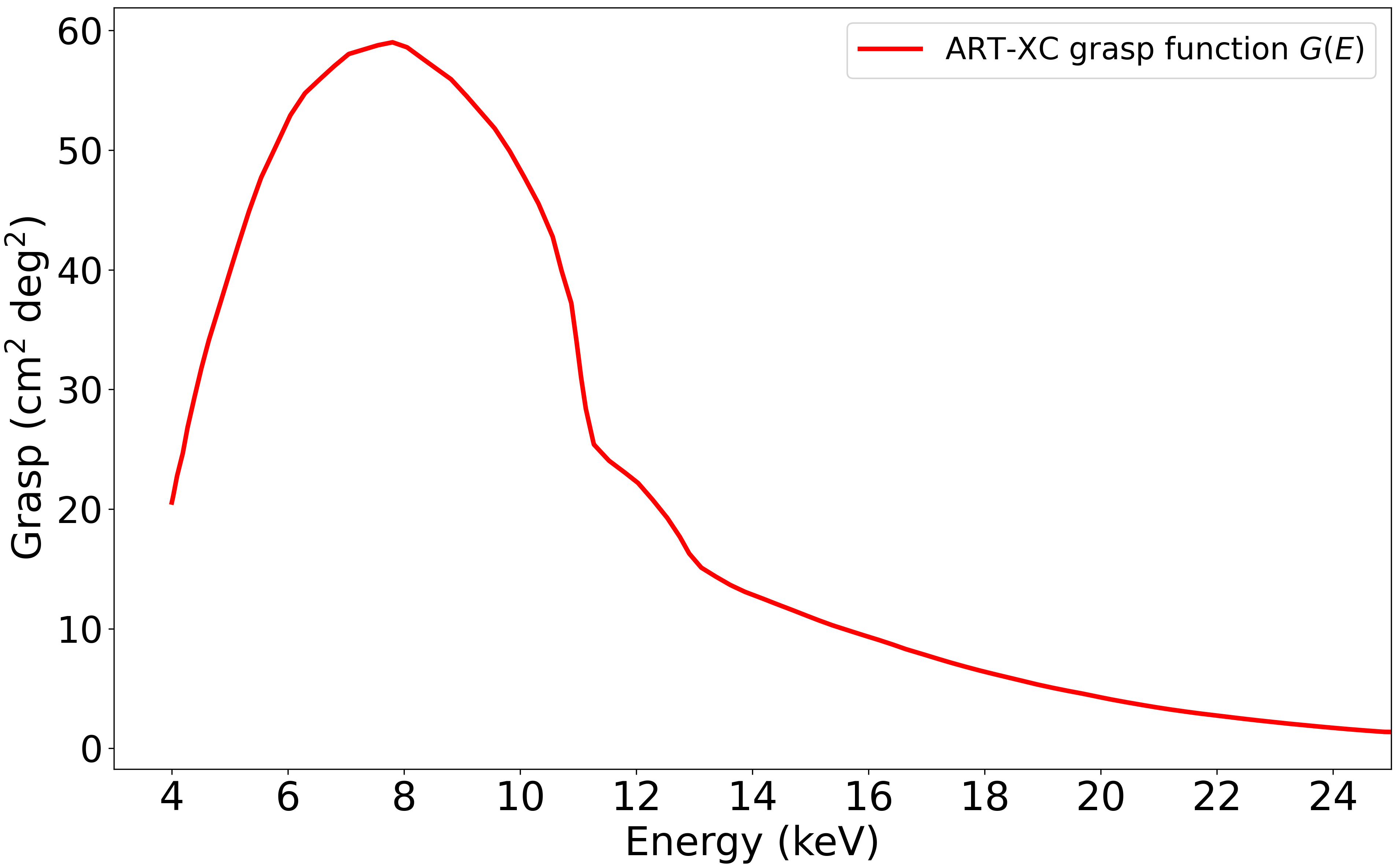}
    \caption{The ART-XC grasp adjusted for zero grade events included in this investigation.}
    \label{fig:grasp}
\end{figure}

The number of signal events is proportional to the signal flux, total observation time of a given region and so-called grasp function $G(E)$. The grasp function is the product of the energy-dependent effective area of the telescope (cm$^2$) and the field of view (deg$^2$). Our grasp function is presented in Fig. \ref{fig:grasp}. 
Two important features of the telescope's mirror system and detectors are  accounted in this grasp function. First, there are two types of photons pass through grazing incidence optical systems; singly reflected and doubly reflected. Doubly reflected photons are used to construct x-ray images in 0.3 deg$^2$ field of view. Singly reflected photons are detected with lower sensitivity in much larger, 2 deg$^2$ field of view. These photons make a significant contribution to the background recorded by the telescope. Since our work examines the x-ray background, we take into account both types of photons. Second, in our work we used only the telescope detector events, that led to the activation of only one detector pixel. Such events (also called zero-grade events) are more likely to be photons than charged particles (see \cite{2019ExA....48..233P}  for details).

The decaying dark matter particles  \eqref{decays} emit photons of energy  $E=m_s/2$ which is registered by detector according to its energy resolution. We approximate ART-XC energy resolution with the Gaussian characterized by full width at half maximum (FWHM) as a function of energy presented in Ref.\,\cite{Pavlinsky:2021ynp}: it is almost constant, FWHM($E$)$\approx 1.2$\,keV with deviations not exceeding 10\%. The number of signal events we expect in a particular bin of energy $E_i$ can be approximated as follows: 
\[
S_i=\frac{A\times 2\sqrt{\log 2}}{\sqrt{\pi} \text{FWHM}(E)}
\exp{\l - \l\frac{ 2(E_i-E)}{\text{FWHM}(E)}\r^2\log{2}\r} \Delta E \,,
\]
with parameter $A$ related to the signal flux as  
\[
\frac{A\Omega^I}{T^I G(E)}=F^I-\lambda F^{II}.
\]
To constrain $A$ at a given energy $E$ (and hence the mixing $\theta$ at a given sterile neutrino mass $m_s$) we construct the log-likelihood function 
\[
l(D|A,E) = \sum_{i=1}^{N^b} \frac{(D_i-B-S_i(A,E))^2}{2\sigma_i^2},
\]
where $\sigma_i^2=\l\Delta D_i\r^2$, i.e. the main errors are pure statistical. To find the upper limit for the signal amplitude $A$, we perform the following procedure; for each energy we increase $A$ to make the following equality true 
\[
-2 \times \l l(D|A,E) - l(D|\hat{A},E) \r = 2.71,
\]
where $\hat{A}$ is the maximum likelihood estimate of A. In the large count limit, the log-likelihood difference reduces to $\Delta \chi^2$ for a single degree of freedom. In terms of Gaussian standard deviations, one side 95\% C.L. upper limit corresponds to $1.64\sigma$, i.e.\ to $\Delta \chi^2 = 2.71$. Consequently we constrain the sterile-neutrino mixing angle as a function of sterile neutrino mass, the results are presented in Fig.\,\ref{fig:results}. 

{\it 4. Conclusion and discussion.}
Our limit (red line in Fig.\,\ref{fig:results}) is the first ever limit based on the x-ray data with uniform sky coverage. In our 12-40\,keV interval of sterile neutrino masses, similar searches for the traces of DM radiative decay were recently performed with {\nustar} observations in various parts of the sky, excluding the Galactic plane ($|b|<15^{\circ}$) \citep{Roach:2022lgo}. In our  study, we confirm the  upper limits obtained there. 
As compared to Ref.\,\cite{Roach:2022lgo}, our analysis also excludes a new region in [$\sin^2(2\theta$, $m_s$] space, i.e., is the hatched area above red line in Fig.~\ref{fig:results}. We should note that this region has been previously disfavored by the {\nustar} observation of the Galactic bulge \cite{2020PhRvD.101j3011R}. Taking the uncertainties of determination of the mass of dark matter in the central regions of the galaxy into account \cite{Barinov:2020hiq}, we expect that our estimates, based on all-sky data are more robust. The main result of our work is that we independently confirm the {\nustar}  results obtained in~\citep{Roach:2022lgo,2020PhRvD.101j3011R}. Note that it is important to confirm these constraints with different experiments, since all 95\% upper limits presented in Fig.\,\ref{fig:results} are inherently noisy due to statistical fluctuations in real data.

The further prospects of ART-XC in exploring the sterile neutrino DM are related to the use of ART-XC data obtained during its Galactic plane survey, undertaken in 2022--2023, with more than a year of total exposure. Also we plan to use the next 2-year ART-XC all-sky survey, which is planned to be started in 2023. We expect that these data will significantly improve our current ART-XC constraints on the decay of the sterile neutrino DM, presented in this paper.

The other important SRG dataset, are the data of eROSITA survey, which should allow for exploring sterile neutrino DM in softer, 0.5--8\,keV energy range, while in 4-8\,keV energy range, covered by the both SRG telescopes, the joint analysis of the data from both ART-XC and eROSITA is possible. The analysis of the fluxes from nearby galaxy clusters (e.g.\ Coma) and analysis of the contribution of unresolved x-ray sources along the lines of Ref.\,\cite{Barinov:2022kfp} can additionally strengthen the inferred constraint on the model parameters of the sterile neutrino DM given in this paper.       

\acknowledgements

{\it Acknowledgements.} The work is supported by the RSF Grant No. 22-12-00271.

\addcontentsline{toc}{chapter}{\bibname}
\bibliographystyle{apsrev4-1}
\bibliography{refs}

\end{document}